\documentclass[12pt]{article}
\usepackage{mathrsfs}
\newsavebox\foobox
\newlength{\foodim}

\usepackage{color}
\usepackage{amsfonts,amssymb,amsmath}
\usepackage[export]{adjustbox}
 
\usepackage{graphicx}
\usepackage[T1]{fontenc}
\usepackage[numbers,sort&compress]{natbib}
\graphicspath{ {./images/} } \textheight 9in \textwidth  6.5in
\newcounter{tempeq}
\begin{document}
\title{\textsf{Entanglement-assisted remote energy transfer}}
\author{B. Mojaveri\thanks{Email: bmojaveri@azaruniv.ac.ir; bmojaveri@gmail.com (corresponding author)}\hspace{.25mm},
\hspace{2mm}R. Jafarzadeh Bahrbeig\thanks{Email:
r.jafarzadeh86@gmail.com}\hspace{.25mm}, \hspace{2mm}M. A. Fasihi
\thanks{Email: a.fasihi@gmail.com} \hspace{.35mm}and\hspace{2mm}N. Abdi
\thanks{Email: nasrinabdi5657@gmail.com}\\\\
{\small {Department of Physics, Azarbaijan Shahid Madani University,
PO Box 51745-406, Tabriz, Iran \,}}} \maketitle
\begin{abstract}
Currently, remote energy transfer and immunity to dissipation are
hot topics in quantum batteries (QBs). In this work, we propose a
protocol to realize energy transfer between two remote atoms (a
quantum charger and a quantum battery) each coupled to a separate
optical cavity with the cavities connected by a fiber. The cavities
and fiber are coupled to their individual baths. After optimizing
inter-system couplings to achieve an efficient transfer, we uncover
the effect of suppressing dissipation by introducing parity
deformation of the cavities fields. We also prove that the
charger-battery entanglement is a consumable resource for energy
storage: it is initially stored until the charger and battery reach
energy balance, and then subsequently consumed to maintain the
increase in energy stored in the battery. The present scheme is the
first execution of energy transfer to a distant battery assisted by
entanglement, which may help better understand quantum
thermodynamics and open new possibilities toward harnessing
decoherence as a resource to improve the charging performance of
QBs.
\end{abstract}
\section{Introduction}
Recent progresses in quantum thermodynamics have provided a deeper
understanding of small-scale energy transfer, and aroused
significant interest in the study of quantum batteries (QBs). QBs
are quantum mechanical systems that allow one to temporarily store
energy in a controllable manner using the principles of quantum
mechanics \cite{Alicki, Hovh, Rossini, Ando0, Liu}. These minatory
devices are charged either by an external agent, or through a
(in)direct interaction between QBs themselves with another quantum
system which acts as a charger. When the battery-charger interaction
is established, QB transits from a lower energy level to a higher
one and gets fully charged. Unlike the classical counterparts, QBs
can benefit from quantum resources, such as quantum coherence,
quantum steering, quantum entanglement and squeezing \cite{Ficher,
Ghosh, Gao0, Gyhm, Salvia}. The advantage of quantum resources is
that they can boost the battery's charging power, speedup the
charging process, and also improve the work that can be extracted
from the battery under the unitary operations. These advantages make
QBs an alternative solution to energy challenges. Alicki and Fannes
were the first to propose QBs and showed that quantum entanglement
can be utilized to improve the amount of work that can be extracted
from a QB under the unitary operations \cite{Alicki}. Since then, to
seek further for quantum advantages of QBs, different theoretical QB
models have been proposed, ranging from single few-level system to
composite interacting systems such as quantum oscillators
\cite{Farina00, Keck, Zhang, Jie, Kerr}, optical and optomechanical
cavities \cite{Ando0, Fer0, Wang00, Bedar, Dou33, Pir, Mami, Zafar,
Fus}, spin chains \cite{Lib, Ghosh, Zakavati, Dimitris, Downing,
Must, NJP, PRL, Mah, Farin}, also the Su-Schrieffer-Heeger and
Sachdev-Ye-Kitaev models \cite{Kitaev00, Dario00, PT, Dario11}. In
parallel to these theoretical developments, experimental efforts
have also been devoted to implement QB using the superconductors
\cite{Fer0, Zhu00, IBM, SUP}, trapped ions \cite{Lv}, quantum dots
\cite{IBM1}, organic microcavities \cite{Quach23}.

 QBs are, generally, charged either by an external agent, or coherent coupling to another quantum system which
acts as a charger. When the battery-charger coupling is established,
QB transits from a lower energy level to a higher one and gets fully
charged. A powerful charging protocol provides a better charging
performance, which is characterized by its storage energy,
efficiency, power, storage capacity as well as ergotropy (the
maximum extractable energy from the battery through cyclic unitary).
The qubit-based quantum systems are one of the widely studied QB
model, which are charged by the external fields \cite{Zhang, Sasi},
or through a (in)direct interaction with other qubits \cite{Farin1,
Haseli, Moj2025, Kamin1}. Previous studies on such batteries have
shown that, apart from the differences in their charging mechanism,
quantum resources play a significant role in the charging
performance \cite{Saliva00, Allahverdyan, Nimmrichter00, Steer}. For
example, more work can be extracted from $N$ identical copies of a
two-level QB using entangling operations \cite{Alicki}. Entangling
operations can perform better also in the charging power when an
array of $N$ identical qubits are charged collectively \cite{Keck,
Kitaev00, Binderm, Binderc}. However, implementation of QB in the
real world faces two major challenges. The first is decoherence
arising from the unavoidable coupling of the QB to its environment.
Unfortunately, decoherence leads to battery energy leakage to the
environment during the charging process, thereby reducing the
charging performance of QB \cite{Farin1, Maze, Manzo, Shaghaghi00,
Delmonte, Safr, Dou44, Dong00, Moj33, Forn, Monsel, Camp, Salimi,
Dou00, Down, Ling, Barra, San0, Pedro, Carega00}. The second major
challenge is that, as the charger-battery distance increases,
battery charging becomes inefficient. During the last few years,
robust charging protocols using dark states \cite{Dark},
decoherence-free subspaces \cite{Dou11}, electromagnetically-induced
transparency \cite{Ali1, Ali2}, repeated quantum measurements
\cite{Sequ, Measurement}, environment engineering \cite{Morrone,
Kamin1, Moj2025, Squeezing, Segal, Borhan1, China}, inhiring an
auxiliary quantum system \cite{Cata} as well as Floquet engineering
\cite{Floq} have been proposed to overcome decoherence of QB. In
addition to the above robust protocols, several other control
techniques such as feedback control \cite{Mitch, Shao, Ios},
frequency modulation \cite{SH}, Bang-Bang modulation \cite{Franc},
adiabatic state preparation protocol  \cite{Segal, Baris, STRAP,
Fasihi}, Lewis-Riesenfeld invariants method \cite{Ning00}, fast
moving technique \cite{Sci}, convergent iterative algorithm
\cite{Borhan} have been exploited to suppress the decoherence
effects on the open QBs. On the other hand a series of approaches
have been devoted to overcome low charging efficiency in the weak
charger-battery coupling condition. For example, a wireless-charging
QB scheme via coupling the QB and the charger to a memory-keeping
environment has been proposed in \cite{Kamin1}. The advantage of
this environment-mediated charging scheme is that Non-Markovian
features of the charger-battery system, stemming from quantum memory
effects, may be utilized to enhance charging performance of the QB.
It has been shown by us, that even if the wireless charging process
is carried out by mediation of a memory-less environment, parity
deformation of the environmental field introduces a memory source to
the environment thereby transforming the Markovian charging process
to the non-Markovian one, which is in favor of improving the
charging performance \cite{Moj2025}. The remote-charging QB has been
realized recently by mediation of a waveguide environment
\cite{Song0}. It has been found that a better charging performance
can be achieved when two bound states are formed in the energy
spectrum of the entire system consisting of the QB, the charger, and
the environment. Furthermore, emphasizing on the role of
entanglement, a long-distance remote charging scheme has been
recently designed based on LC circuits \cite{LC0}.

 Motivated by these considerations and the challenges outlined above, we
design a novel energy-transfer scheme for a two-qubit QB, in
presence of dissipative environments. In our scheme, one of the
qubits is the QB and the other is the charger which are embedded in
two distant lossy cavities connected by a leaky optical fiber, where
each cavity interacts with a single qubit. We note that the qubits
do not interact with each other. We will examine the impact of
inter-system couplings on the stored energy and extractable work of
the quantum battery, and show that a high efficiency charging is
achieved when the energy storage capacity of the optical fiber is
severely degraded. We will use the parity-deformed quantization of
cavity fields to control the decoherence effect of the environment
on the energy transfer. In fact, parity deformation induces specific
intensity-dependent couplings between the atom and cavity, the
cavity and fiber, as well as the cavity and bath, thus allowing
control of the system dynamics by tuning the deformation parameter.
We will show that in our charging scheme, weakening these couplings
leads to suppression decoherence effect and improvement of charging
performance. We will also discover the relationship between
entanglement and energy stored in the QB, and numerically show that
improvement in energy storage and extractable work requires the
consumption of the initial entanglement, which is generated between
charger and battery before reaching energy equilibrium. The rest of
the paper has been structured as follows. The physical model, open
dynamics of the charger-battery system and an explicit expression of
their evolved reduced density matrix are given in the second
section. Section 3 is devoted to introduce and describe several
figures of merit for characterizing the performance of QBs. Section
4 discuss our results. Finally, section 5 concludes this
paper.\begin{figure} \centering\includegraphics[keepaspectratio,
width=1\textwidth]{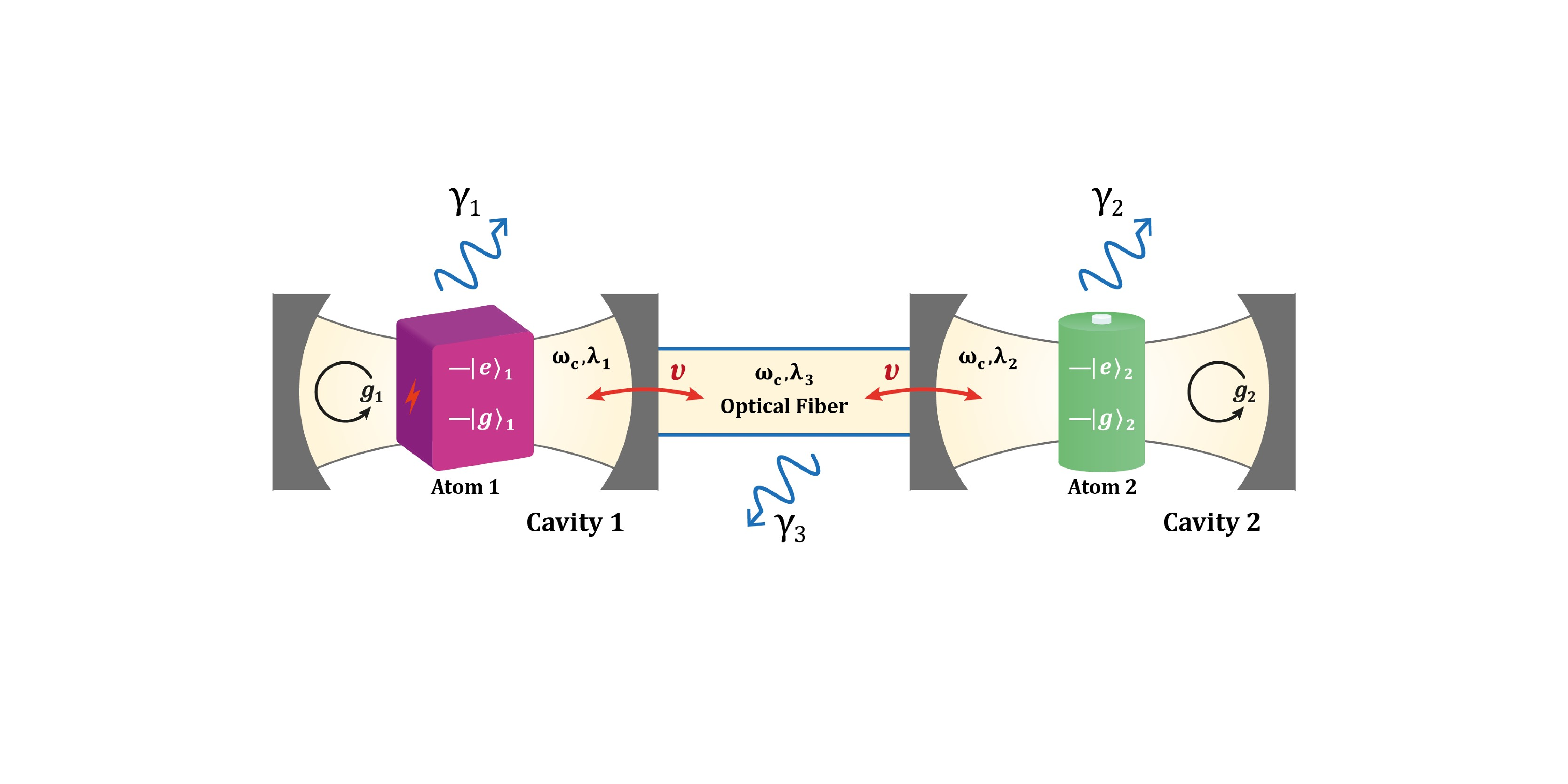} \caption{Schematic illustration of
charging process of a remote battery. The battery and charger qubits
are located separately at two distant cavities, which are connected
by a short-length optical fiber.}
\end{figure}
\section{QB model and its wireless charging process}
The remote-charging QB model that we are going to discuss consists
of two distant two-level atoms, the atom 1 as a charger and the atom
2 as a QB, located separately at two cavities (1 and 2), which are
connected by a short-length optical fiber (an additional single-mode
cavity) (shown in Fig. 1). The cavities and fiber are open and
surrounded by a bosonic reservoir at thermal equilibrium (zero
temperature). For simplicity we consider the short fiber limit
$2l\Gamma/(2\pi c)\ll1$, where $l$ is the length of the fibre and
$\Gamma$ is the decay rate of the cavity modes into a continuum of
fiber modes, only one mode (resonant) of the fiber will interact
with the cavity modes \cite{Serafini}. The cavity electromagnetic
radiation is described by a single-mode parity-deformed field,
characterized by the annihilation and creation generators
$\hat{\mathbf{a}}$ and $\hat{\mathbf{a}}^{\dagger}$ of the Wigner
(para-Bose) Lie algebra. The Wigner algebra \cite{Green0, Green1,
VAS} is a parity-deformed version of the harmonic oscillator
algebra, which is defined by the commutation relations on the
annihilation and creation generators $\hat{\mathbf{a}}$,
$\hat{\mathbf{a}}^{\dagger}$, and a parity operator $\hat{R}$:
\renewcommand\theequation{\arabic{tempeq}\alph{equation}}
\setcounter{equation}{-1} \addtocounter{tempeq}{1}
\begin{eqnarray}\label{Wigner}
&&\hspace{-14mm}\left[\hat{\mathbf{a}},
\hat{\mathbf{a}}^{\dagger}\right]=1+2\nu\hat{R},\,\,\,\left\{\hat{R},
\hat{\mathbf{a}}\right\}=\left\{\hat{R},
\hat{\mathbf{a}}^{\dagger}\right\}=0.
\end{eqnarray}
where the real constant $\nu\in (-0.5, \infty)$ denotes the parity
deformation parameter. Note that in the limit $\nu\rightarrow 0$, we
will lose the anti-commutation relations of Eq. (\ref{Wigner}) so
that the remaining commutation relation returns the harmonic
oscillator algebra. Using a nonlinear map \cite{Poly, Curt} the
generators $\hat{\mathbf{a}}$, $\hat{\mathbf{a}}^{\dagger}$ and
$\hat{R}$ has been realized in terms of the well-known photon
annihilation and creation operators $\hat{a}$ and
$\hat{a}^{\dagger}$ through the noncanonical transformations
\cite{Dask0}\renewcommand\theequation{\arabic{tempeq}\alph{equation}}
\setcounter{equation}{-1} \addtocounter{tempeq}{1}
\begin{eqnarray}\label{NL}
&&\hspace{-14mm}\hat{\mathbf{a}}=\hat{a}F(\hat{n}), \qquad\qquad
\hat{\mathbf{a}}^{\dagger}=F(\hat{n})\hat{a}^{\dagger}, \qquad\qquad
\hat{R}=(-1)^{\hat{n}},
\end{eqnarray}
where $\hat{n}=\hat{a}^{\dagger}\hat{a}$ is the photon number
operator, and $F(\hat{n})$ is an analytic function of the photon
number operator and is equal to
$F(\hat{n})=\bigg(1+\frac{\nu}{\hat{n}}(1-(-1)^{\hat{n}})\bigg)^{\frac{1}{2}}$.
As a consequence of the noncanonical transformations (\ref{NL}), the
para-Bose algebra can be interpreted as a nonlinear oscillator
algebra \cite{Manko} which is represented with the photon-number
basis $|n\rangle$, as below \cite{Bashir0, Mojt}
\renewcommand\theequation{\arabic{tempeq}\alph{equation}}
\setcounter{equation}{0} \addtocounter{tempeq}{1}
\begin{eqnarray}
&&\hspace{-14mm}\hat{\mathbf{a}}\left|2n\right\rangle=\sqrt{2n}\left|2n-1\right\rangle,\quad\quad\,\quad\quad\quad\quad\,
\hat{\mathbf{a}}\left|2n+1\right\rangle=\sqrt{2n+2\nu+1}\left|2n\right\rangle,\label{Ladering1}\\
&&\hspace{-14mm}\hat{\mathbf{a}}^{\dag}\left|2n\right\rangle=\sqrt{2n+2\nu+1}
\left|2n+1\right\rangle,\quad\quad \hat{\mathbf{a}}^{\dag}\left|2n+1\right\rangle=\sqrt{2n+2}\left|2n+2\right\rangle,\label{Ladering2}\\
&&\hspace{-14.5mm} \hat{R}\left|n\right\rangle=
(-1)^n\left|n\right\rangle.
\end{eqnarray}

 The total Hamiltonian of the QB model is (hereafter, we set
$\hbar=1$):
\renewcommand\theequation{\arabic{tempeq}\alph{equation}}
\setcounter{equation}{-1}
\addtocounter{tempeq}{1}\begin{eqnarray}\label{TootalH}
&&\hspace{-12cm}\hat{H}=\hat{H}_S+\hat{H}_{Bath}+\hat{H}_{int},\label{Ho}
\end{eqnarray}
where $\hat{H}_{S}$ is the Hamiltonian of the QB system forms of the
atoms, cavities, and fiber; $\hat{H}_{Bath}$ is the bath
Hamiltonian, and $\hat{H}_{int}$ describe, the interaction of the QB
system with the bath. The Hamiltonian $\hat{H}_S$ is given by
\renewcommand\theequation{\arabic{tempeq}\alph{equation}}
\setcounter{equation}{-1}\addtocounter{tempeq}{1}\begin{eqnarray}\label{HS}
&&\hspace{-14mm}\hat{H}_S=\sum_{i=1,2}\hat{H}_{\nu_i}+\frac{\omega_c}{2}\{\hat{a}^{\dagger}_3,\hat{a}_3\}
+\lambda\hat{a}_3\big(\hat{\mathbf{a}}^{\dagger}_1+\hat{\mathbf{a}}^{\dagger}_2\big)+
\lambda\hat{a}^{\dagger}_3\big(\hat{\mathbf{a}}_1+\hat{\mathbf{a}}_2\big),
\end{eqnarray}
where
$\hat{H}_{\nu_i}=\frac{\omega_a}{2}\hat{\sigma}^{(i)}_{z}+\frac{\omega_c}{2}\{\hat{\mathbf{a}}^{\dagger}_i,\hat{\mathbf{a}}_i\}+
\eta\big(\hat{\sigma}^{(i)}_{-}\hat{\mathbf{a}}^{\dagger}_i+\hat{\sigma}^{(i)}_{+}\hat{\mathbf{a}}_i\big)$
is the Hamiltonian of the parity-deformed Jaynes-Cummings model
describing the interaction of the parity-deformed field mode of
cavity $i$, characterized by frequency $\omega_c$ and lowering
(raising) operators $\hat{\mathbf{a}}_i$
($\hat{\mathbf{a}}^{\dagger}_i$), with atom $i$, characterized by
transition frequency $\omega_a$ and lowering (raising) operators
$\hat{\sigma}^{(i)}_{-}=|g\rangle_i\,_i\langle e|$
($\hat{\sigma}^{(i)}_{+}=|e\rangle_i\,_i\langle g|$), with a real
coupling strength $\eta$. Moreover, the second term
$\frac{\omega_c}{2}\{\hat{a}^{\dagger}_3,\hat{a}_3\}$ corresponds
the free Hamiltonian of the fiber, and the last terms
$\lambda\hat{a}_3\big(\hat{\mathbf{a}}^{\dagger}_1+\hat{\mathbf{a}}^{\dagger}_2\big)+
\lambda\hat{a}^{\dagger}_3\big(\hat{\mathbf{a}}_1+\hat{\mathbf{a}}_2\big)$
describe the cavity-fiber interaction in the RWA. Here $\hat{a}_3$
($\hat{a}^{\dagger}_3$) is lowering (raising) boson operator for the
fibre, $\omega_c$ is the fiber frequency which is the same as that
of the cavity frequencies, and $\lambda$ denotes the cavity-fiber
hopping strength. In the photon number basis $|n\rangle$,
Hamiltonian (\ref{HS}) has the form
\renewcommand\theequation{\arabic{tempeq}\alph{equation}}
\setcounter{equation}{-1} \addtocounter{tempeq}{1}
\begin{eqnarray}\label{NI}
&&\hspace{-1.2cm}
\hat{H}_S=\omega_c\hat{a}^{\dagger}_3\hat{a}_3+\sum_{i=1,2}\bigg[\frac{\omega_a}{2}\hat{\sigma}^{(i)}_{z}+\omega_c\hat{a}^{\dagger}_i\hat{a}_i
+\eta\big(F(\hat{n}_i)\hat{\sigma}^{(i)}_{-}\hat{a}^{\dagger}_i+H.C.\big)+\lambda\big(F(\hat{n}_i)\hat{a}_3\hat{a}^{\dagger}_i+H.C.\big)\bigg]
 \label{f-d}
\end{eqnarray}
describing a composite system consisting of two identical qubits,
which are trapped individually in two distant single-mode bosonic
cavities coupled by an optical fiber, were the atom-cavity as well
as cavity-fiber interactions are intensity dependent with the
intensity function
$F(\hat{n}_i)=\bigg(1+\frac{\nu_i}{\hat{n}_i}(1-(-1)^{\hat{n}_i})\bigg)^{\frac{1}{2}}$.

 Furthermore, by indicating with $b_{ik}(b^{\dagger}_{ik})$ the bath bosonic
annihilation (creation) operators for the cavity 1-fiber-cavity 2
array ($i=1,2,3$), the bath and interaction Hamiltonians appearing
in Eq. (\ref{TootalH}) are, respectively,
\renewcommand\theequation{\arabic{tempeq}\alph{equation}}
\setcounter{equation}{0} \addtocounter{tempeq}{1}\begin{eqnarray}
&&\hspace{-14mm}\hat{H}_{Bath}=\sum_{k=0}^{\infty}\sum_{i=1}^3\omega_{ik}\hat{b}^{\dagger}_{ik}\hat{b}_{ik},\\
&&\hspace{-14mm}\hat{H}_{int}=\sum_{k=0}^{\infty}\sum_{i=1,2}\bigg[g_{ik}\big(\hat{a}_i
F(\hat{n}_i)
+F(\hat{n}_i)\hat{a}^{\dagger}_i\big)\big(\hat{b}_{ik}+\hat{b}^{\dagger}_{ik}\big)\bigg]+\sum_{k=0}^{\infty}
g_{3k}\big(\hat{a}_3+\hat{a}^{\dagger}_3\big)\big(\hat{b}_{3k}+\hat{b}^{\dagger}_{3k}\big),\label{HINT}
\end{eqnarray}
where $g_{ik}$ are the system-bath coupling constants. Notice that,
for simplicity, we take into account only the dissipative
interaction between the system and the bath, and neglect the pure
dephasing type environmental noises. In addition, we neglect the
dissipation corresponding to the spontaneous emission of atoms
because this kind of dissipation is insignificant in the cavity QED
regime considered in our model.

 In the following, we begin to calculate the open dynamical
evolution of the QB system involving qubits, cavities and fiber. Due
to the fact that in optical cavities thermal noise is negligible, we
are here interested in the dynamics of QB system at zero
temperature. We assume that at most a single excitation is allowed
into the QB system, and therefore the \textit{bare basis} of the
system are: $|1\rangle=|eg000\rangle$, $ |2\rangle=|gg100\rangle$,
$|3\rangle=|gg001\rangle$, $|4\rangle=|gg010\rangle$,
$|5\rangle=|ge000\rangle$, and $|6\rangle=|gg000\rangle$. Note that,
in this notation, the order of each subsystem state in the system's
basis is $|\textrm{qubit} 1- \textrm{qubit}\, 2- \textrm{cavity}\,
1- \textrm{cavity}\, 2- \textrm{fiber}\rangle$. Using these basis,
the Hamiltonian $\hat{H}_S$ in equation (\ref{HS}) after ignoring
the constant term $\omega_c(\nu_1+\nu_2+3/2)$ can be expressed as
the following matrix form
\renewcommand\theequation{\arabic{tempeq}\alph{equation}}
\setcounter{equation}{-1} \addtocounter{tempeq}{1}\begin{eqnarray}
\hat{H}_{S}=\left(
              \begin{array}{cccccc}
                0 & \eta\sqrt{2\nu_1+1} & 0 & 0 & 0 & 0 \\
                \eta\sqrt{2\nu_1+1} & \Delta& \lambda\sqrt{(2\nu_1+1)} & 0 & 0 & 0 \\
                0 & \lambda\sqrt{(2\nu_1+1)} & \Delta & \lambda\sqrt{(2\nu_2+1)} & 0 & 0 \\
                0 & 0 & \lambda\sqrt{(2\nu_2+1)} & \Delta&  \eta\sqrt{2\nu_2+1} & 0 \\
                0 & 0 & 0 &  \eta\sqrt{2\nu_2+1} & 0 & 0 \\
                0 & 0 & 0 & 0 & 0 & -\omega_a \\
              \end{array}
            \right),
\end{eqnarray}
where $\Delta=\omega_c-\omega_a$. In the following, for simplicity,
we focus on the case $\nu_1=\nu_2=\nu$. In that case, the
eigenvalues of the Hamiltonian $H_S$ are
\renewcommand\theequation{\arabic{tempeq}\alph{equation}}
\setcounter{equation}{-1}\addtocounter{tempeq}{1}
\begin{eqnarray}\label{MasterA}
&&\hspace{-14mm}E_n=\frac{2}{3}\sqrt{a_2^2-3a_1}\cos\left[\frac{\theta+2(n-1)\pi}{3}\right]-\frac{a_2}{3},\,\,\,\,\,(n=1,2,3),\nonumber\\
&&\hspace{-14mm}E_{4(5)}=\frac{1}{2}\bigg[a_2-\Delta\pm\sqrt{\frac{4a_0}{\Delta}+(a_2-\Delta)^2}\bigg],\,\,\,\,\,E_6=-\omega_a,
\end{eqnarray}
where
\renewcommand\theequation{\arabic{tempeq}\alph{equation}}
\setcounter{equation}{-1}\addtocounter{tempeq}{1}
\begin{eqnarray}\label{MasterA}
&&\hspace{-14mm}a_0=\eta^2\Delta(2\nu+1),\,\,\,a_1=-(\eta^2+2\lambda^2)(2\nu+1)+\Delta^2+2\nu\Delta\omega_c,\nonumber\\
&&\hspace{-14mm}a_2=2(\Delta+\nu\omega_c),\,\,\,\theta=\cos^{-1}\left[\frac{r}{\sqrt{q^3}}\right],
\end{eqnarray}
with
\renewcommand\theequation{\arabic{tempeq}\alph{equation}}
\setcounter{equation}{-1}\addtocounter{tempeq}{1}
\begin{eqnarray}\label{MasterA}
&&\hspace{-14mm}r=\frac{9a_1a_2-27a_0-2a_2^3}{54},\,\,\,q=\frac{a_2^2-3a_1}{9}.
\end{eqnarray}
Also, the corresponding eigenstates (\textit{dressed states}) of the
Hamiltonian $\hat{H}_S$ are
\renewcommand\theequation{\arabic{tempeq}\alph{equation}}
\setcounter{equation}{-1}\addtocounter{tempeq}{1}
\begin{eqnarray}\label{EigeS}
&&\hspace{-14mm}|\Phi_n\rangle=\sum_{k=1}^5 c_{nk}|k\rangle,\,\,
(n=1,2,.., 5),\,\,\,|\Phi_6\rangle=|6\rangle,
\end{eqnarray}
where the nonzero coefficients $c_{nm}$ are
\renewcommand\theequation{\arabic{tempeq}\alph{equation}}
\setcounter{equation}{-1}\addtocounter{tempeq}{1}
\begin{eqnarray}\label{coffi}
&&\hspace{-16mm}c_{11}=c_{21}=c_{31}=-c_{41}=-c_{51}=1,\,c_{n2}=c_{n4}=\frac{E_n}{\eta\sqrt{2\nu+1}},\nonumber\\
&&\hspace{-16mm}c_{n3}=-\frac{\eta^2(2\nu+1) +\lambda
E_n(\Delta-E_n+2\nu\omega_c)}{\eta\lambda(2\nu+1)}, c_{n5}=1,
\,\,\,(n=1,2,3).
\end{eqnarray}
Taking into account the notations $\vec{\Phi}=(|\Phi_1\rangle,
|\Phi_2\rangle,..., |\Phi_6\rangle)^{T}$ and
$\vec{J}=(|1\rangle,|2\rangle, ..., |6\rangle)^{T}$, the equation
(\ref{EigeS}) can be rewritten with a matric form, as follows
\renewcommand\theequation{\arabic{tempeq}\alph{equation}}
\setcounter{equation}{-1}\addtocounter{tempeq}{1}
\begin{eqnarray}\label{Vector}
&&\hspace{-14mm}\vec{\Phi}=C \vec{J},
\end{eqnarray}
where $C$ is a $6\times 6$ matrix with the nonzero components
$c_{nm}$ given in (\ref{coffi}).
\subsection{The open dynamics of the QB}
Let us assume that the cavities and fiber are interacting weakly
with the bath. In this case, under the Born-Markov and secular
approximations the dynamics of the open QB system involving qubits,
cavities and fiber is described via the following Markovian master
equation \cite{Breuer0}
\renewcommand\theequation{\arabic{tempeq}\alph{equation}}
\setcounter{equation}{-1}\addtocounter{tempeq}{1}
\begin{eqnarray}\label{MasterA}
&&\hspace{-3cm}\dot{\hat{\rho}}(t)=-i[\hat{H}_S,\hat{\rho}(t)]+\sum_{i=1}^3\gamma_i\left[\hat{A}_{i}(\Omega)\,\hat{\rho}(t)\hat{A}_{i}^{\dag}(\Omega)
-\frac{1}{2}\{\hat{A}_{i}^{\dag}(\Omega)\hat{A}_{i}(\Omega),\hat{\rho}(t)\}\right],
\end{eqnarray}
where $\hat{\rho}(t)$ is reduced density matrix of the QB system,
$\gamma_i$ ($i=1,2,3$) denote the photon decay rates from the
cavities and fiber, respectively, and $\hat{A}_{i}(\Omega)$ are the
transition operators for the corresponding photonic subsystems. The
transition operators are constructed as follows
\renewcommand\theequation{\arabic{tempeq}\alph{equation}}
\setcounter{equation}{0}\addtocounter{tempeq}{1}
\begin{eqnarray}\label{Jump}
&&\hspace{-3cm}\hat{A}_{1(2)}(\Omega)=\sum_{\Omega_{ij}=\Omega>0}\left|\Phi_i\right\rangle\left\langle
\Phi_i\right|\left(\hat{\mathbf{a}}_{1(2)}+\hat{\mathbf{a}}^{\dagger}_{1(2)}\right)\left|\Phi_j\right\rangle\left\langle
\Phi_j\right|,\\
&&\hspace{-3cm}\hat{A}_{3}(\Omega)=\sum_{\Omega_{ij}=\Omega>0}\left|\Phi_i\right\rangle\left\langle
\Phi_i\right|\left(\hat{a}_3+\hat{a}^{\dagger}_3\right)\left|\Phi_j\right\rangle\left\langle
\Phi_j\right|,
\end{eqnarray} for all positive eigenfrequencies $\Omega = \Omega_{ij}=E_i-E_j>0$, corresponding to the transitions
$|\Phi_i\rangle\rightarrow|\Phi_j\rangle$. Using the laddering
relations (\ref{Ladering1}) and (\ref{Ladering2}), we can derive an
explicit expression for the transition operators
\renewcommand\theequation{\arabic{tempeq}\alph{equation}}
\setcounter{equation}{0}\addtocounter{tempeq}{1}
\begin{eqnarray}\label{Jump1}
&&\hspace{-3cm}\hat{A}_{1}(\Omega)=\sum_{n=1}^{5}c_{n2}\sqrt{2\nu_1+1}\left|\Phi_6\right\rangle\left\langle
\Phi_n\right|,\\
&&\hspace{-3cm}\hat{A}_{2}(\Omega)=\sum_{n=1}^{5}c_{n4}\sqrt{2\nu_2+1}\left|\Phi_6\right\rangle\left\langle
\Phi_n\right|,\\
&&\hspace{-3cm}\hat{A}_{3}(\Omega)=\sum_{n=1}^{5}c_{3n}\left|\Phi_6\right\rangle\left\langle
\Phi_n\right|.
\end{eqnarray}
To solve the master equation (\ref{MasterA}) we need to recast it in
the eigenvector basis $|\Phi_j\rangle$, because this simplifies the
master equation into two sets of decoupled differential equations,
one set for the population (diagonal) elements and the other set for
the coherence (non-diagonal) elements. The differential equations
for the coherence and population elements of the density matrix
$\hat{\rho}(t)$ are:
\renewcommand\theequation{\arabic{tempeq}\alph{equation}}
\setcounter{equation}{0}\addtocounter{tempeq}{1}
\begin{eqnarray}\label{Dequ}
&&\hspace{-3cm}\dot{\rho}_{nn}(t)=-\gamma_{nn}\rho_{nn}(t),\quad\quad \dot{\rho}_{66}(t)=\sum_{n=1}^5\gamma_{nn}\rho_{nn}(t),\\
&&\hspace{-3cm}\dot{\rho}_{nm}(t)=\bigg[-i\Omega_{nm}-\frac{\gamma_{nn}+\gamma_{mm}}{2}\bigg]\rho_{nm}(t),\quad\quad
\dot{\rho}_{6n}(t)=0,
\end{eqnarray}
where
$\gamma_{nn}=\gamma_1(2\nu_1+1)|c_{n2}|^2+\gamma_2(2\nu_2+1)|c_{n4}|^2+\gamma_3|c_{n3}|^2$.
The time dependence of the population and coherence elements can be
now found easily as
\renewcommand\theequation{\arabic{tempeq}\alph{equation}}
\setcounter{equation}{0}\addtocounter{tempeq}{1}
\begin{eqnarray}\label{Delement}
&&\hspace{-3cm}\rho_{nn}(t)=\rho_{nn}(0)e^{-\gamma_{nn}t},\quad\quad \rho_{66}(t)=\sum_{n=1}^5\rho_{nn}(0)[1-e^{\gamma_{nn}t}],\\
&&\hspace{-3cm}\rho_{nm}(t)=\rho_{nm}(0)e^{-\frac{1}{2}\left[2i\Omega_{nm}+\gamma_{nn}+\gamma_{mm}\right]t},\quad\quad
\rho_{6n}(t)=\rho_{6n}(0).
\end{eqnarray}
The resulting expressions serve as a convenient starting point for
analyzing the charging performance of the QB. In fact, after
choosing the initial state of the QB system $\hat{\rho}(0)$ and the
values of the various parameters, they enable us to compute the
reduced density matrix and then energetics of battery's qubit (the
qubit 2) at time $t$.
\section{Charging performance of the battery}
We now investigate stored energy and ergotropy which are two figures
of merit enabling quantitative evaluation of the charging
performance by examining the effect of inter-system couplings and
parity deformation of the cavities fields on the charging
performance. In particular, we show how parity deformation of the
cavities fields can lead to a robust energy transfer.

  The energy of the battery at time $t$ is represented by the expectation value of
the battery's Hamiltonian
$\hat{H}_B=\frac{\omega_a}{2}\hat{\sigma}^{(2)}_{z}$, with respect
to the reduced density matrix $\hat{\rho}_{B}(t)$, as
$E_B(t)=\texttt{Tr}\left[\hat{\rho}_B(t)\hat{ H}_B\right]$. Hence,
by taking into account the initial state of the battery as
$|eg000\rangle$, the change in battery energy during charging
process can be measured by
\renewcommand\theequation{\arabic{tempeq}\alph{equation}}
\setcounter{equation}{-1} \addtocounter{tempeq}{1}\begin{equation}
\Delta E_B=E_B(t)-E_B(0)=\omega_a\tilde{\rho}_{55}(t),
\end{equation}
where $\tilde{\rho}_{55}(t)$ is the battery state in the bare basis,
which can be computed using the basis transformation relation
$\tilde{\rho}(t)=C^{-1}\hat{\rho}(t)C$, as follows
\renewcommand\theequation{\arabic{tempeq}\alph{equation}}
\setcounter{equation}{-1}\addtocounter{tempeq}{1}
\begin{eqnarray}\label{EB}
&&\hspace{-3cm}\tilde{\rho}_{55}(t)=\hat{\rho}_{33}=\sum_{n=1}^{5}\rho_{nn}(t)|c_{n5}|^2+2\sum_{n\neq
m=1}^{5}\mathfrak{Re} \left[\rho_{nm}(t)\bar{c}_{m5}c_{n5}\right],
\end{eqnarray}
with $\mathfrak{Re}$ and $\bar{z}$ indicating, respectively, the
real part and complex conjugate of a complex variable. Similarly, by
considering the charger's Hamiltonian
$\hat{H}_C=\frac{\omega_a}{2}\hat{\sigma}^{(1)}_{z}$ as well as
charger's density matrix $\hat{\rho}_{C}(t)$, reduction of internal
energy of charger during at time $t$ can be analytically quantified
by
\renewcommand\theequation{\arabic{tempeq}\alph{equation}}
\setcounter{equation}{-1} \addtocounter{tempeq}{1}\begin{equation}
E_{C}(t)=\omega_a\tilde{\rho}_{11}(t),
\end{equation}
where $\tilde{\rho}_{11}(t)$ is given by
\renewcommand\theequation{\arabic{tempeq}\alph{equation}}
\setcounter{equation}{-1}\addtocounter{tempeq}{1}
\begin{eqnarray}\label{EC}
&&\hspace{-3cm}\tilde{\rho}_{11}(t)=\hat{\rho}_{22}=\sum_{n=1}^{5}\rho_{nn}(t)|c_{n1}|^2+2\sum_{n\neq
m=1}^{5}\mathfrak{Re} \left[\rho_{nm}(t)\bar{c}_{m1}c_{n1}\right].
\end{eqnarray}
We note that often not all stored energy in the QB may be
extractable as work, Ergotropy is the maximum amount of the energy
that can be extracted from the battery through unitary protocols,
and therefore is crucial indicator to judge the performance of the
battery. According to \cite{Allahverdyan}, the ergotropy of a given
$\rho$ with respect to a reference Hamiltonian $H$ is defined as
follows
\renewcommand\theequation{\arabic{tempeq}\alph{equation}}
\setcounter{equation}{-1}
\addtocounter{tempeq}{1}\begin{eqnarray}\label{Ergo}
&&\hspace{-3cm}\mathcal{W}=\texttt{Tr}\{\rho
H\}-\texttt{min}_{U}\,\texttt{Tr}\{U\rho
U^{\dagger}H\},\label{ergotropy}
\end{eqnarray}
where minimization is taken over all unitary transformations $U$
acting locally on the reduced density matrix $\rho$. If we consider
the spectral decomposition of $H$ and $\rho$, i.e.
$H=\sum_{i}\varepsilon_{i}|\varepsilon_{i}\rangle\langle\varepsilon_{i}|$
and $\rho=r_{i}|r_{i}\rangle\langle r_{i}|$ ordered as
$\varepsilon_{i}\leq\varepsilon_{i+1}$, and $r_{i+1}\leq r_{i}$ then
the optimal unitary cycle gets
$U=\sum_{k}|\varepsilon_{k}\rangle\langle r_{k}|$. Therefore,
according to Eq. (\ref{Ergo}), the ergotropy corresponding to the
battery state $\tilde{\rho}_{55}(t)$ can be obtained as
\renewcommand\theequation{\arabic{tempeq}\alph{equation}}
\setcounter{equation}{-1}
\addtocounter{tempeq}{1}\begin{equation}\label{Ergo1}
\mathcal{W}_B=\omega_a\left(2\tilde{\rho}_{55}(t)-1\right)\Theta
\left(\tilde{\rho}_{55}(t)-\frac{1}{2}\right),
\end{equation}
where $\Theta(x-x_0)$ is the Heaviside function, which satisfies
$\Theta(x-x_0)=0$ for $x<x_0$, $\Theta(x-x_0)=\frac{1}{2}$ for
$x=x_0$ and $\Theta(x-x_0)=1$ for $x>x_0$. According to Eq.
(\ref{EB}) and (\ref{Ergo1}), it is no difficult to prove that the
energy can be extracted from the battery through unitary protocols,
only when the battery is charged to more than half its
capacity.\begin{figure}\centering\includegraphics[keepaspectratio,width=1.01\textwidth]{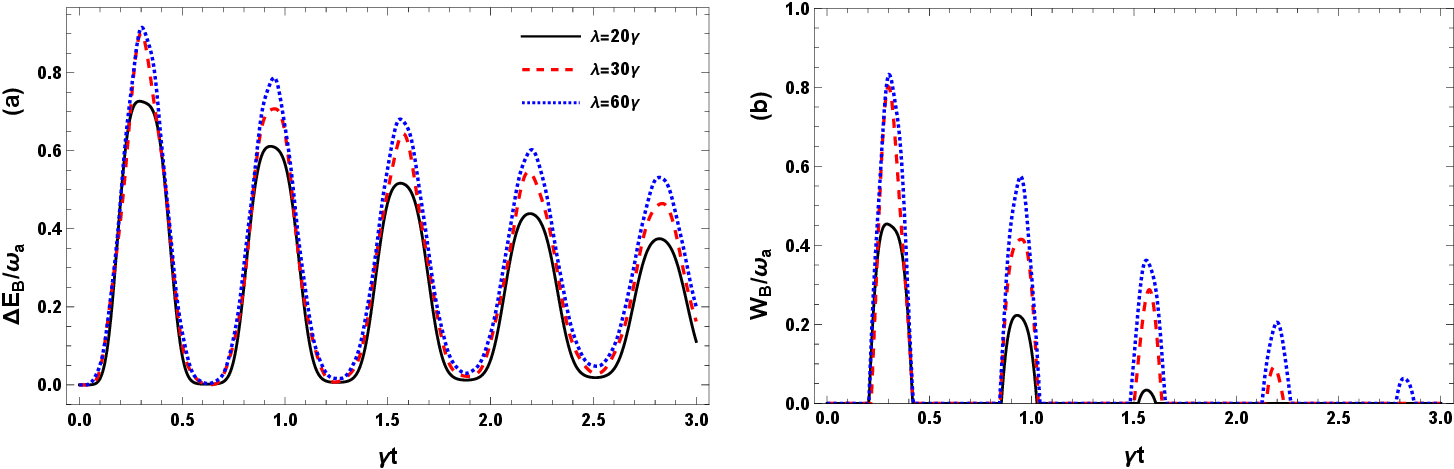}
\caption{Dynamics of (a) the normalized stored energy $\Delta
E_B(t)/\omega_a$ and (b) normalized ergotropy
$\mathcal{W}_B/\omega_a$ for the different values of $\lambda$ by
setting $\eta=10\gamma$ and $\nu=0$.}
\end{figure}

 In the following, we analyze dynamical behavior of stored energy as well as ergotropy, and examine the impact of the
inter-system couplings, i.e. the qubit-cavity and cavity-fiber
couplings on the charging performance with emphasis on the control
role of parity deformation of cavities. For this purpose, we
consider an initial state $|1\rangle=|e,g,000\rangle$, in which the
battery's qubit, cavities and fiber are in their ground state while
the charger's qubit is in its excited state. For the sake of
simplicity, we consider the resonant case (i.e. $\Delta=0$), set
$\gamma_1=\gamma_2=\gamma_3=\gamma$ to have same damping rate, and
normalize the rest of system parameters by the damping rate
$\gamma=10$ GHz. We also choose a mid-infrared (MIR) transition
frequency i.e., $\omega_a=800\pi\gamma$ for our two-level atoms. The
normalized values of coupling constants will be tuned to satisfy the
approximations on the Markovian master equation i.e.,
$\eta/\gamma\gg 1 $ and $\lambda/ \gamma\gg 1$.

 In Fig. 2, we plot (a) the stored energy $\Delta E_B$ and (b) ergotropy as a function of
the dimensionless quantity $\gamma t$ for different values of the
cavity-fiber hooping constant $\lambda$ by choosing $\nu=0$, where
the cavities' electromagnetic radiation is described by a
single-mode bosonic field. According to panel (a), the stored energy
exhibit an oscillatory damping behavior. This behavior is a common
characteristic of open QBs, which indicates that during the charging
process, energy leakages into the bath due to decoherence effect
arising in the system. In panel (a), we observe that increasing the
fiber-cavity hooping strength $\lambda$, increases the amplitude of
energy oscillations without significantly affecting their period,
implying that stronger fiber-cavity hooping facilitates a robust
charging process. Indeed, as can be seen in panel (b), such an
increase in amplitude of stored energy, respectively, will lead to
increase of the time region during which a considerable amount of
stored energy can be extracted through cyclic unitary. These
findings underscore the constructive role of cavity-fiber hooping in
boosting the charging performance of the
QB.\begin{figure}\centering\includegraphics[keepaspectratio,width=1.01\textwidth]{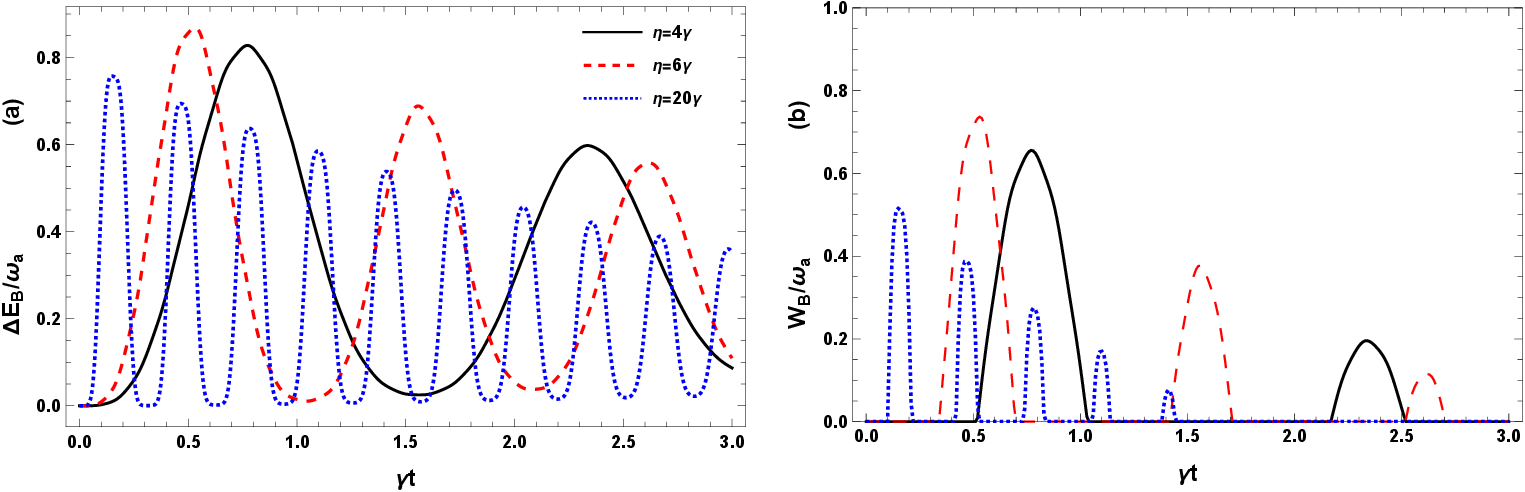}
\caption{Dynamics of (a) the normalized stored energy $\Delta
E_B(t)/\omega_a$ and (b) normalized ergotropy
$\mathcal{W}_B/\omega_a$ for the different values of $\eta$ by
setting $\lambda=40\gamma$ and $\nu=0$.}\end{figure}

 In Fig. 3 we present the impact of atom-cavity coupling on the
charging performance of the QB. Oscillatory damping behavior of the
normalized stored energy $\Delta E_B$ and ergotropy for different
values of the atom-cavity coupling $\eta$ by setting $\nu=0$ are
depicted in Figs. 3(a) and 3(b), respectively. It is clear that, in
the case of $\eta\ll \lambda$, gradual growth of atom-cavity
coupling strength not only increases the maximum of the stored
energy, but also accelerates the charging process, facilitating
faster energy transfer. However, further increasing $\eta$ beyond
the $\eta\ll \lambda$ leads to a gradual decline in the maximum of
the stored energy. A similar qualitative impact of $\eta$ is
observed on the ergotropy, as expected. As can be seen in Fig. 3(b),
the maximum ergotropy increases regularly by gradual growth of the
qubit-cavity coupling strength, and extraction of ergotropy happens
sooner the larger $\eta$ is.

 In order to gain a physical perspective on the features of stored energy and ergotropy in Figs. 2 and 3,
we need to deepen our understanding of how energy is transferred
between the battery and the charger, and also explore the role of
inter-system couplings in suppressing decoherence of QB. To respond
to this need, in Figs. 4 (5) we display dynamics of
$\tilde{\rho}_{ii}$ ($i=1,2,..,6$), as a function of the
dimensionless quantity $\gamma t$ for the parameters corresponding
to Fig. 2 and (3). We note the populations $\tilde{\rho}_{11}$,
$\tilde{\rho}_{55}$ and $\tilde{\rho}_{66}$ correspond the
normalized internal energy of charger, stored energy in the battery
$\Delta E_B/\omega_a$, and the loss in energy due to dissipation,
respectively, while $\tilde{\rho}_{22}$, $\tilde{\rho}_{44}$ and
$\tilde{\rho}_{33}$ correspond the normalized internal energy of the
cavity 1, 2 and the fiber, respectively.\begin{figure}
\centering\includegraphics[keepaspectratio,width=1\textwidth]{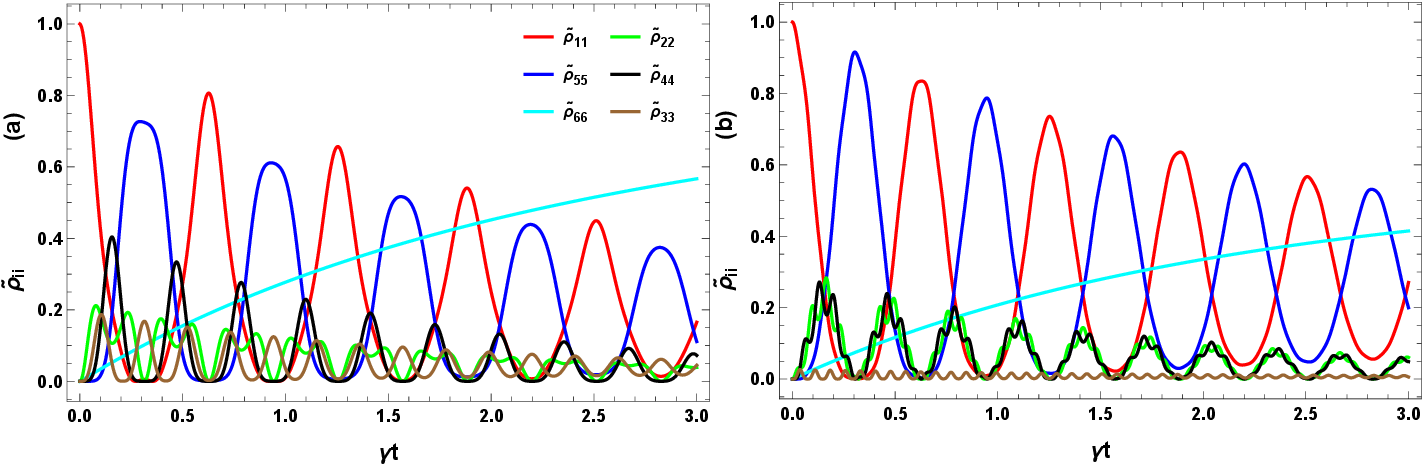}
\caption{Dynamics of population distribution $\tilde{\rho}_{ii}$ for
(a) $\lambda=20\gamma$ and (b) $\lambda=60\gamma$ by setting
$\eta=10\gamma$ and $\nu=0$.}
\end{figure}
\begin{figure} \centering \includegraphics[keepaspectratio,
width=1\textwidth]{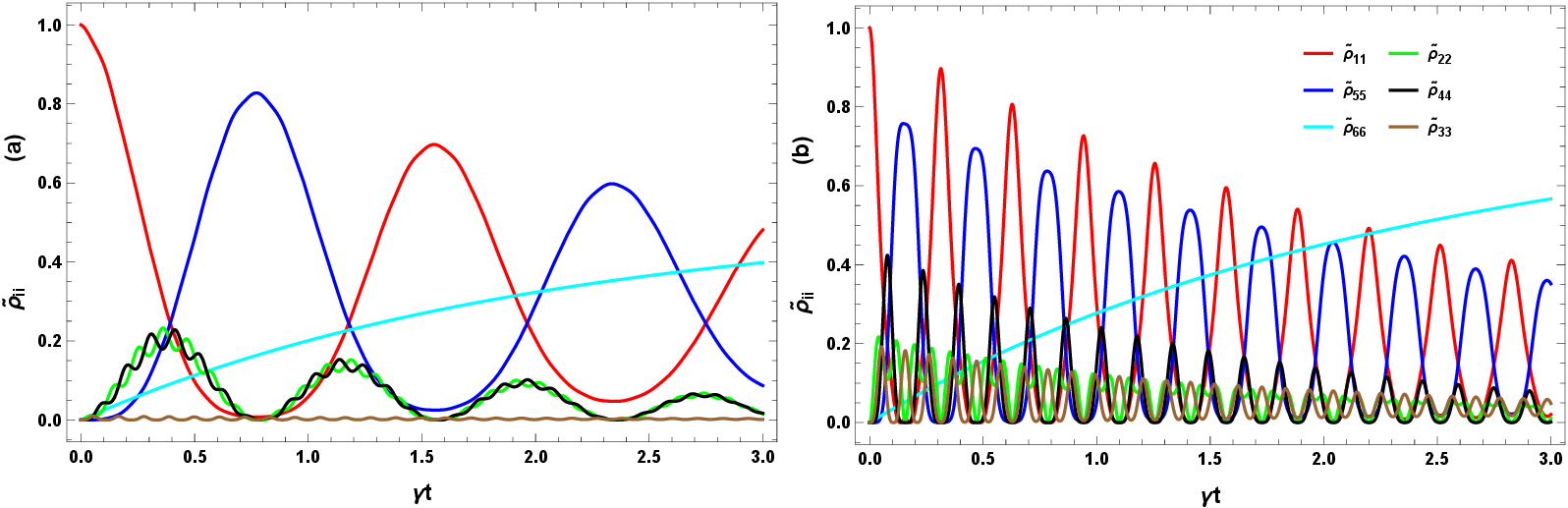} \caption{Dynamics of population
distribution $\tilde{\rho}_{ii}$ for (a) $\eta=4\gamma$ and (b)
$\eta=20\gamma$ by setting $\lambda=40\gamma$ and $\nu=0$.}
\end{figure}

 These figures represent the energy transfer from the charger qubit to the battery qubit and vice versa, in detail
and reveal the influence of $\lambda$ and $\eta$ on the
energy-transfer performance. What is generally shown in these
figures, is that immediately as soon as the interaction is switched
on, the charger qubit starts sharing its initial excitation (energy)
in turn with the cavity 1, fiber and cavity 2. Then, over time the
$\tilde{\rho}_{22}$, $\tilde{\rho}_{33}$ and $\tilde{\rho}_{44}$
come in and out of phase with $\tilde{\rho}_{11}$ or
$\tilde{\rho}_{55}$ and beat together to transfer energy from the
charger to the battery and vice versa. Fig. 4(a) shows that in the
case where the cavity-fiber coupling strength is near to the
atom-cavity coupling (e.g. $\lambda\simeq 2\eta$),
$\tilde{\rho}_{22}$, $\tilde{\rho}_{33}$ and $\tilde{\rho}_{44}$ get
out of phase from $\tilde{\rho}_{11}$ and $\tilde{\rho}_{55}$. As a
result, a significant amount of the excitation is trapped in the
charging mediator (i.e. in the cavities and fiber) during the
charging stages, which favors a reduction in the maximum energy
stored in QB as well as an increase in the decoherence rate.
However, as can be seen in Fig. 4(b), when the cavity-fiber coupling
is far from the atom-cavity coupling ($\lambda\gg \eta$), fiber does
not consume almost any energy, hence the energy dissipation from the
charging mediator is greatly suppressed, which leads to a robust
energy transfer. On the other hand, during a charging stage, as
$\lambda$ and $\eta$ get far from each other, $\tilde{\rho}_{22}$
and $\tilde{\rho}_{44}$ become in phase with $\tilde{\rho}_{55}$
until battery reaches energy equilibrium with the charger, but then
get out of phase until the battery reaches its peak. In turn, it
suggests an increase in the maximum stored energy of the QB, given
that no energy is accumulated in the charging mediator.

 Furthermore, it is also evident by comparing Figs. 5(a) and 5(b), as
$\eta$ gets closer to $\lambda$ in the $\lambda\gg \eta$ regime,
both energy dissipation rate and energy-transfer rate are amplified.
This behavior can be expected because stronger inter-system
couplings tends to make qubits more intertwined to accelerate the
excitation transfer. In addition, based on the above discussions as
$\eta$ gets closer to $\lambda$, $\tilde{\rho}_{22}$ and
$\tilde{\rho}_{44}$ interfere destructively in favor of increasing
the energy
dissipation.\begin{figure}\centering\includegraphics[keepaspectratio,width=1.01\textwidth]{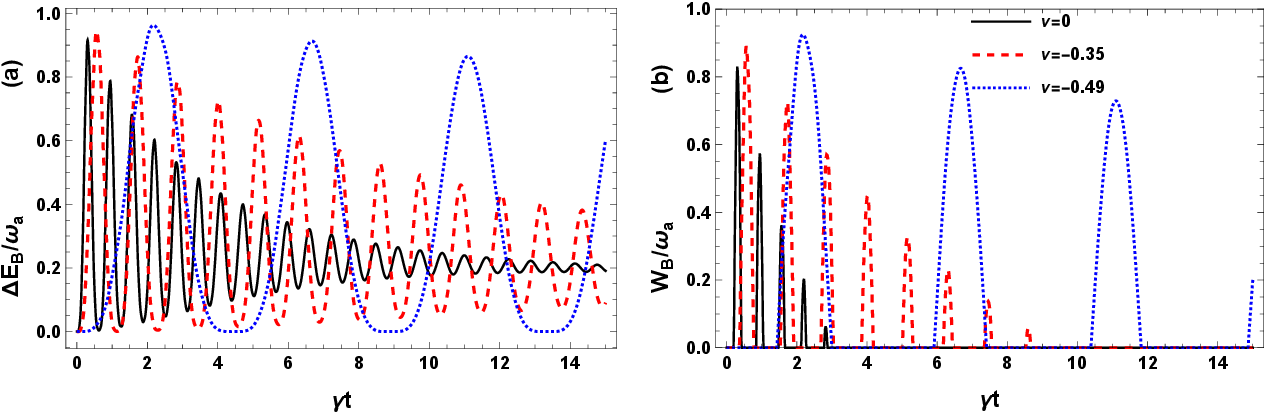}
\caption{Dynamics of (a) the normalized stored energy $\Delta
E_B(t)/\omega_a$ and (b) normalized ergotropy
$\mathcal{W}_B/\omega_a$ for the different values of $\nu$ by
setting $\lambda=60\gamma$ and $\eta=10\gamma$.}
\end{figure}

 Now, we examine the effect of parity deformation of the cavity field on the charging performance. The
results for the stored energy and ergotropy are illustrated in Figs.
6 (a) and 6(b), respectively. In the absence of the parity
deformation, $\Delta E_B(t)/\omega_a$ and $\mathcal{W}_B/\omega_a$
exhibit oscillatory behavior with rapid damping, indicating that the
energy dissipation in the charging process is high. However, when
parity deformation is introduced, the energy dissipation is
significantly reduced, as very strong oscillatory behavior is
observed in panels (a) and (b) for negative values of the parity
deformation parameter $\nu$. Our results show that a more robust
energy transfer is achieved when the deformation parameter $\nu$ is
adjusted near its lower bound.

 In order to understand how parity deformation can suppress the energy dissipation of the QB,
we turn to Eqs. (\ref{NI}) and (\ref{HINT}), where we showed that
parity deformation of cavities fields makes the inter-system
couplings in the Hamiltonian $\hat{H}_S$ as well as the system-bath
coupling in Hamiltonian $\hat{H}_{int}$ dependent on $\nu$. It is
clear that as $\nu$ approaches to its minimum value, the system-bath
coupling is weakened, thus the energy dissipation is significantly
suppressed. On the other hand, by regularly increasing the negative
values of $\nu$, the strength of inter-system couplings decreases,
but their ratio remains unchanged, so the energy is transferred with
delay.
\section{Relationship between battery-charger coherency and energy transfer}
Now, we analyze the relationship between the battery-charger
entanglement and energy storage during the charging process. To
quantify entanglement, we use here the so-called concurrence
measure.

 Performing a trace over the cavities and fiber
degrees of freedom of the density matrix $\tilde{\rho}(t)$, will
ultimately leave a reduced density matrix of the charger-battery
system which is represented in the bare state basis $\{|ee\rangle,
|e,g\rangle, |g,e\rangle, |gg\rangle\}$, as follows
\renewcommand\theequation{\arabic{tempeq}\alph{equation}}
\setcounter{equation}{-1}\addtocounter{tempeq}{1}
\begin{eqnarray}\label{DM}
&&\hspace{-1cm}\hat{\rho}_{CB}(t)=\left(
                  \begin{array}{cccc}
                    0 & 0 & 0 & 0\\
      0 & \tilde{\rho}_{11} & \tilde{\rho}_{15} & 0 \\
      0 & \tilde{\rho}_{51} & \tilde{\rho}_{55} & 0 \\
      0 & 0 & 0 & 1-\tilde{\rho}_{11}-\tilde{\rho}_{55}
                  \end{array}
                \right).\end{eqnarray}

\begin{figure} \centering\includegraphics[keepaspectratio, width=1.01\textwidth]{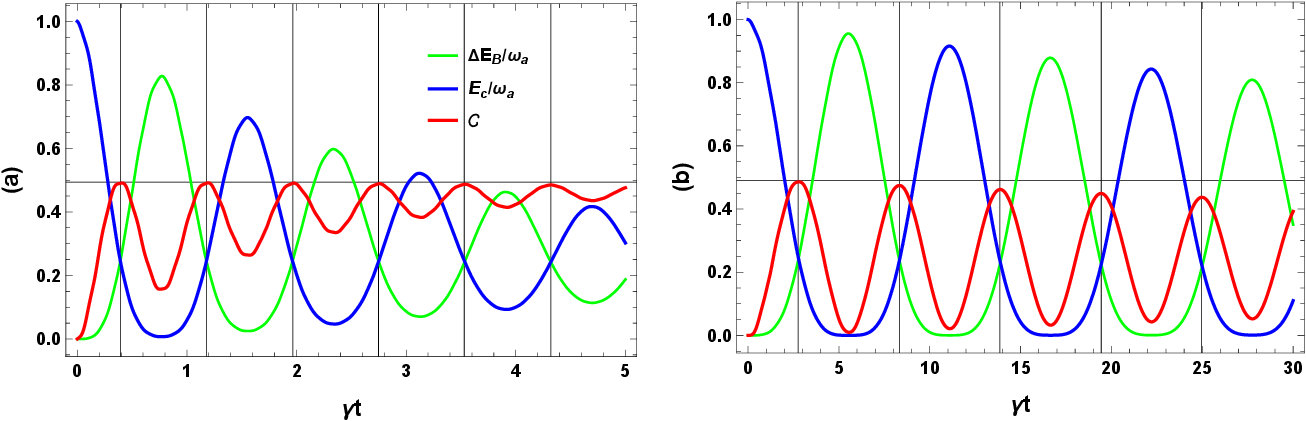}
\caption{Dynamics of (a) the normalized stored energy $\Delta
E_B(t)/\omega_a$, internal energy of charger $E_C(t)/\omega_a$ and
concurrence $\mathcal{C}$ for (a) $\nu=0$ and (b) $\nu=-0.49$ by
setting $\lambda=40\gamma$ and $\eta=4\gamma$.}
\end{figure}
For the above battery-charger density matrix, which is an example of
the so-called X state, the concurrence has the expression
$\mathcal{C}=2\max\{0,\,|\tilde{\rho}_{15}|\}$.

 Fig. 7, illustrates the dynamics of $\Delta
E_B(t)/\omega_a$, $E_C(t)/\omega_a$ and $\mathcal{C}$ in the
$\lambda\gg \eta$ regime, where no energy is trapped in the charging
mediator and thus a powerful charging performance is achieved. Fig.
7(a) focuses on illustrating the relationship of battery-charger
entanglement in the absence of parity deformation. It can be seen
that the maximum entanglement occurs only when the charger and
battery have the same energy. As soon as either deviates from this
energy balance, concurrence is suppressed regardless of the energy
transfer direction. This suggests that the initial transfer of
energy requires the accumulation of entanglement, while higher
energy storage comes at the cost of consuming some of accumulated
entanglement. In this perspective, battery-charger entanglement can
be considered as an essential consumable resource for energy
transfer. Fig. 7(b) focuses on illustrating the relationship of
battery-charger entanglement when the cavities field are parity
deformed. It can be concluded the relationship does also hold in the
presence of parity deformation. That is, after reaching energy
balance, energy transfer continues at the expense of battery-charger
entanglement consumption until the stored energy reaches its peak.
Comparing 7(a) and (b) reveals the fact that parity deformation
increases the tendency to consume entanglement in favor of
suppressing the decoherence in the QB.
\section{Outlook and summary}
In summary, we proposed a mechanism to realize large-distance energy
transfer between two remote qubits in a cavity quantum
electrodynamics (CQED) network with losses. The qubits are located
separately at two distant single-mode bosonic cavities, which are
connected by an optical fiber. The cavities and fiber are also
surrounded by a zero-temperature bosonic bath. The qubit-cavity as
well as cavity-bath couplings were considered intensity dependent,
where the intensity function is taking into account as
$F(\hat{n}_i)=\bigg(1+\frac{\nu_i}{\hat{n}_i}(1-(-1)^{\hat{n}_i})\bigg)^{\frac{1}{2}}$.
These specific intensity-dependent couplings are realized if the
cavity annihilation and creation operators are considered
parity-deformed. We studied the effect of inter-system parameters on
the charging performance, and showed that by careful tailoring
inter-system couplings, a powerful charging performance could be
achieved. We showed that parity-deformation of cavity fields are
essential for realizing a high-energy charging process. In addition,
we discovered the relationship between entanglement and the energy
transfer and found that charger-battery entanglement serves as an
essential resource for enhancing energy storage: entanglement is
accumulated between the charger and battery until they reach energy
balance, then subsequently consumed until the maximum energy storage
and ergotropy is achieved. The role of parity deformation in this
relationship is to significantly increase the consumption of
entanglement in favor of enhancing stored energy and ergotropy. Our
findings suggest that parity deformation of cavity photons could be
a novel interesting solution to overcome the decoherence effect in a
noisy cavity-QED network which provides a robust energy transfer to
a remote battery. Our protocol could shedding up a new light on how
entanglement can be used to more accurate monitoring and predicting
of the charging status of the QB.\\\\
\textbf{\large{Data availability}}\\ The datasets used and analysed
during the current study available from the corresponding author on
reasonable request.

\end{document}